\documentclass[conference]{IEEEtran}

\usepackage[letterpaper, top=0.5in, bottom=0.55in, left=0.75in, right=0.75in]{geometry}
\usepackage{tikz}
\usepackage{pgfplots}
\usepgfplotslibrary{external}
\usepackage{amsmath}
\usepackage{amsfonts}
\usepackage{graphicx}
\usepackage{amssymb}
\usepackage{amstext}
\usepackage{latexsym}
\usepackage{array,tabularx}
\usepackage{subcaption}
\usepackage{mathtools}
\usepackage{url}
\usepackage{comment}
\usepackage{dirtytalk}
\usepackage[ruled,vlined]{algorithm2e}
\usepackage{cite}
\usetikzlibrary{matrix,decorations.pathreplacing}

\DeclareMathOperator*{\minimize}{minimize}
 
\DeclareMathOperator*{\argmin}{arg\,min}

\newtheorem{theorem}{Theorem}

\newtheorem{lemma}{Lemma}

\newtheorem{definition}{Definition}
\newtheorem{corollary}{Corollary}

\newcommand{\interior}[1]{%
  {\kern0pt#1}^{\mathrm{o}}%
}

\begin{document}

\title{Heterogeneous Differential Privacy via Graphs}

\author{Sahel Torkamani\IEEEauthorrefmark{1}, Javad B. Ebrahimi\IEEEauthorrefmark{1}\IEEEauthorrefmark{4}, Parastoo Sadeghi\IEEEauthorrefmark{2}, Rafael G. L. D'Oliveira\IEEEauthorrefmark{3}, Muriel Médard\IEEEauthorrefmark{3} 
\\ \IEEEauthorrefmark{1} Sharif University of Technology, Tehran, Iran, \{sahel.torkamani, javad.ebrahimi\}@sharif.edu\\
% change order of emails
 \IEEEauthorrefmark{4}Institute for Research in Fundamental Sciences (IPM), Tehran, Iran\\
 \IEEEauthorrefmark{2}SEIT, University of New South Wales, Canberra, Australia, p.sadeghi@unsw.edu.au\\
\IEEEauthorrefmark{3}RLE, Massachusetts Institute of Technology, USA, \{rafaeld, medard\}@mit.edu\\   
}
\maketitle

\begin{abstract}
We generalize a previous framework for designing utility-optimal differentially private (DP) mechanisms via graphs, where datasets are vertices in the graph and edges represent dataset neighborhood. The boundary set contains datasets where an individual's response changes the binary-valued query compared to its neighbors. Previous work was limited to the homogeneous case where the privacy parameter $\varepsilon$ across all datasets was the same and the mechanism at boundary datasets was identical. In our work, the mechanism can take different distributions at the boundary and the privacy parameter $\varepsilon$ is a function of neighboring datasets, which recovers an earlier definition of personalized DP as special case. The problem is how to extend the mechanism, which is only defined at the boundary set, to other datasets in the graph in a computationally efficient and utility optimal manner. Using the concept of strongest induced DP condition we solve this problem efficiently in polynomial time (in the size of the graph).
\end{abstract}

%\emph{This paper is eligible for the Jack Keil Wolf ISIT Student Paper Award.}

\section{Introduction}

Differential privacy (DP) \cite{dwork2006calibrating} is a mathematical standard for quantifying the privacy performance of a data publishing or data analysis mechanism \cite{survey_2017}. To conceal the presence of any individual in the dataset, DP mechanisms perturb the query response or the outcome of an analysis according to a random distribution.   The main DP parameter is called $\varepsilon$.  If $\varepsilon$ is small, then any mechanism output is almost as likely to occur whether or not any particular individual's data was used in the database. 

Despite many scientific and operational challenges \cite{Censusissues}, the United States Census Bureau has implemented differential privacy for the 2020 Census release \cite{USCBadopts}. One challenge, which is also documented in many other works including \cite{epsilonvote, economic}, is the difficulty in choosing an appropriate value for $\varepsilon$. Two possible reasons for such a challenge are as follows.

First, differential privacy is not well-equipped with theories that maximize utility subject to a privacy constraint or minimize $\varepsilon$ subject to a utility constraint \cite{Censusissues}. In \cite{staircase, ghosh2012universally, natashalaplace}, the staircase, geometric and Laplace mechanisms were respectively identified as utility-maximizing mechanisms under various notions of utility. However, only the global sensitivity of the query across all datasets is taken into account. Such \emph{data-independent} mechanisms can adversely affect utility, especially when an individual's response does not change the query outcome compared to any of its neighboring datasets \cite{individual}. \emph{Data-dependent} mechanisms aim to enhance utility. However, since utility is not provably optimized, the challenge remains to determine which data-dependent algorithm is best for a given application \cite{data_depenedent}.

Second, a ``one-size-fits-all" \cite{personalized} approach to setting a global privacy level can be damaging to both utility and privacy. Current implementations of differential privacy lack sufficient flexibility for accommodating \emph{data-dependent} privacy setting. For example, there may be minority groups whose data must be better protected. There may also be statutory mandates, demanding publication of certain datasets with more accuracy. The authors of \cite{epsilonvote, personalized} present several social reasons in favor of incorporating users' preferences when choosing $\varepsilon$.

 Towards addressing these challenges, the authors in \cite{RafnoDP2colorPaper} proposed a methodology for data-dependent utility-optimal mechanism design for binary-valued queries. This was done via representing datasets as vertices and dataset neighborhoods as edges on a graph. \emph{Boundary} datasets are those where an individual's data changes the query outcome compared to its neighbors. For the case that the mechanism was defined only \emph{partially} at the boundary datasets, \cite{RafnoDP2colorPaper} showed it is possible to \emph{extend} the mechanism over the entire graph in an optimal manner. To solve the problem efficiently, \cite{RafnoDP2colorPaper} focused on the \emph{homogeneous} case where the partial mechanism had the same probability distribution at the boundary and also $\varepsilon$ was the same across the graph. However, an efficient solution to the general problem remained open.

This paper generalizes the work \cite{RafnoDP2colorPaper} in two main directions.\footnote{We remark that \cite{RafnoDP2colorPaper} considered approximate $(\varepsilon, \delta)$-DP. Here we set $\delta = 0$ and consider pure-DP. This will make the analysis manageable.} First, we study \emph{heterogeneous mechanisms} where the partial mechanism can have different probability distributions at the boundary. Second, we study a \emph{general heterogeneous privacy setting} on neighboring datasets, which recovers personalized DP \cite{personalized} as a special case. Efficiently solving both generalizations required a radically different way of thinking about the problem compared to \cite{RafnoDP2colorPaper}. Instead of using graph morphism to simple path graphs, we use the partial mechanism as \emph{seed to optimally grow} via the concept of \emph{strongest induced DP condition}. We show this can be done in polynomial time. 

After recalling standard definitions for graphs and differential privacy in Section \ref{sec:standard}, we introduce heterogeneous DP in Section \ref{HDP}. Section \ref{sec:main} presents our main results in a semi-informal manner, focusing on insights and intuitions. Section \ref{sec:technical} contains the technical statements and the Algorithm for finding the optimal mechanism. %Proofs can be found in \cite{ourpaper}.

\section{Differential Privacy and Utility via Graphs}\label{sec:standard}
Let $\mathcal{G}(V,E)$ be a simple, connected, and undirected graph with vertex set $V$ and edge set $E$. A sequence of vertices $u = u_0, u_1, \cdots, u_n = v$ is said to form a path from $u$ to $v$, denoted by $(u,v)$-path, if $(u_0, u_1), \cdots, (u_{n-1},u_n) \in E$.

\begin{definition}[$(u,v)$-path set]
For every two vertices $u,v \in V$, we define $\mathcal{P}(u,v) := \{\text{all the } (u,v)\text{-paths in } \mathcal{G}\}$.
For subsets $A_1,A_2 \subseteq V$, we define $\mathcal{P}(A_1,A_2):= \cup _{u \in A_1, v \in A_2} \mathcal{P}(u,v)$. Finally, $\mathcal{P}$ is the set of all paths in $\mathcal{G}$.
\end{definition}

Let $\rho$ be a path and $w$ be a neighbor of $\rho$'s tail not on $\rho$. The path obtained from adding $w$ to $\rho$ is denoted by $\rho w$.

\begin{definition}[Neighborhood]
The neighborhood of a subset $S \subseteq V$ of the vertices, denoted by $N(S)$ is the set of all the vertices in $V\setminus S$ which are connected to at least one element of $S$ by an edge.
\end{definition}

In this work, the vertices represent datasets and the edges represent neighborhood relationships between pairs of datasets. Neighboring datasets $u$ and $v$ are also denoted by $u\sim v$. The true query function $T: V\to Q$, associates to each dataset a query value from a finite set $Q$.

\begin{definition}[Boundary set]\label{def:boundary}
The \emph{boundary} set of $\mathcal G$ with respect to $T$ is denoted by $\partial _T (\mathcal{G})$ and is the set of vertices in $\mathcal G$ whose neighborhood contains at least one vertex with a different true query value. Formally, 
\begin{align*}
\partial _T (\mathcal{G}) = \{u \in V: \exists v \in N(u), T(v) \neq T(u)\}.
\end{align*}
\end{definition}

For privately responding to $T$, a privacy-preserving mechanism $\mathcal{M}$ randomizes the response. 

\begin{definition}[Differential privacy \cite{dwork2006calibrating}]
	\label{def:dp} Let $\varepsilon \geq 0$. Then, a mechanism $\mathcal{M}:V \to Q$ on $\mathcal{G}$ is $\varepsilon$-differentially private (in short is $\varepsilon$-DP) if, for every $u \sim v$ and $\mathcal{S} \subseteq Q$,  \[ \Pr [\mathcal{M}(u) \in \mathcal{S}] \leq e^\varepsilon \Pr [\mathcal{M}(v) \in \mathcal{S}].\]
	\end{definition}

In this paper we consider the case where $Q = \{1, 2\}$ of binary-valued queries. It then suffices to use the following notion of \emph{binary-valued} differential privacy.
\begin{definition}[Binary-valued differential privacy]\label{def:binaryDP}
	Let $\varepsilon \geq 0$ and $p : V \to [0,1]$. We say $p$ is binary-valued $\varepsilon$-DP if for every pair $u \sim v$, we have:
	\begin{align} \label{eq: p1}
	p(u) &\leq e^\varepsilon p(v), \\
	1-p(v) &\leq e^\varepsilon (1-p(u)).\label{eq: p2}
	\end{align}
	\end{definition}
Due to the symmetry of dataset neighborhood, $v \sim u$ will yield the other two inequalities involving $p(v)$ and $p(u)$.
\begin{lemma}
Let $\mathcal{M}: V \to \{1,2\}$ be a mechanism. Then, $\mathcal{M}$ is $\varepsilon$-DP if $p:=\Pr[\mathcal{M}(v)=1]$ is binary-valued $\varepsilon$-DP. 
\end{lemma}
% proof sentence
The proof is straightforward and is omitted here. The following definition captures the optimal utility of a binary mechanism over the space of datasets. Roughly speaking, a mechanism is said to be optimal if for every dataset $v \in V$, the probability of correctly outputting the true query value $T(v)$ is the highest it can be. Recall that $p := \Pr[\mathcal{M} (v) =1]$ and $1-p := \Pr[\mathcal{M} (v) = 2]$.
\begin{definition}[Optimal binary mechanism]\label{def:optimal}
 A binary-valued $\varepsilon$-DP mechanism $p^*$ is said to be optimal on $\mathcal{G}$ if, for every other binary-valued $\varepsilon$-DP mechanism $p$ on $\mathcal{G}$ and every vertex $v \in V$, we have:
\begin{align}
    \begin{cases}
     p^*(v) \geq  p(v), \quad &\text{if }  T(v) = 1,\\ 
     1-p^*(v) \geq  1-p(v), \quad &\text{if }  T(v) = 2.
    \end{cases}
\end{align}
We denote this ordering with respect to $T$ by $p \leq_T p^*$.
\end{definition}

\section{Heterogeneous Differential Privacy}\label{HDP}
The standard definition of differential privacy in Definition~\ref{def:dp} is \emph{homogeneous} in the sense that the privacy conditions between any two neighboring datasets are given by the same $\varepsilon$. In this section, we extend the homogeneous $\varepsilon$-DP to the heterogeneous case. For each edge $(u,v) \in E$, there is an $\varepsilon(u,v)$, which specifies how neighboring datasets $u$ and $v$ should be protected with respect to each other.  Throughout the paper, we assume $\varepsilon(\cdot)$ is symmetric, i.e, $\varepsilon(u,v) = \varepsilon(v, u)$, for every $(u,v) \in E$.

\begin{definition}[Heterogeneous differential privacy]
	\label{def:hetDP} A mechanism $\mathcal{M}:V \to Q$ on $\mathcal{G}$ is heterogeneous $\varepsilon(\cdot)$-differentially private if, for every $u \sim v$ and $\mathcal{S} \subseteq Q$, \[ \Pr [\mathcal{M}(u) \in \mathcal{S}] \leq e^{\varepsilon(u,v)} \Pr [\mathcal{M}(v) \in \mathcal{S}].\]
	\end{definition}
If $\varepsilon(\cdot) = \varepsilon$ is a constant function, we recover the standard differential privacy in Definition \ref{def:dp}. A small $\varepsilon(u,v)$ means high privacy and vice versa. Whenever we write $\varepsilon(\cdot)$-DP as opposed to just $\varepsilon$-DP, it is to emphasize that we mean a heterogeneous DP mechanism on $\mathcal{G}$. 

Definition \ref{def:hetDP} is more general than the personalized DP \cite{personalized} as follows. In \cite{personalized}, the variability of $\varepsilon$ is tied to the identity of an individual $i$. To clarify, for any $v = (v^1, \cdots, v^i, \cdots, v^n) \in V$, where $n$ is the dimension of the dataset, let $V^{-i}(v) \subset V$ denote all vertices in $V$ whose $i$-th element differs from $v$. That is, $v^{-i}= (v^1, \cdots, v'^i, \cdots, v^n)$. In \cite{personalized}, $\varepsilon(v,v^{-i}) = \varepsilon_i$ for all $v^{-i} \in V^{-i}$. In this paper, there is no such constraint and $\varepsilon(v, v^{-i})$ has full degrees of freedom to depend on both $v$ and $v^{-i}$. 

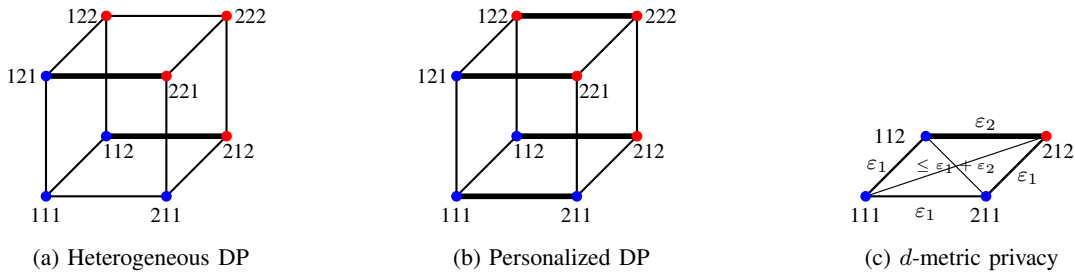
\begin{figure*}[!t]
\begin{subfigure}[t]{0.30\textwidth}
    \centering
    \begin{tikzpicture}[scale=0.8]

\draw[color=black, thick] (1,1) -- (1,3);
\draw[color=black, thick] (2,2) -- (2,4);
\draw[color=black, thick] (1,1) -- (3,1);
\draw[color=black, line width=0.75mm] (2,2) -- (4,2);

\draw[color=black, thick] (3,1) -- (3,3);
\draw[color=black, line width=0.75mm] (1,3) -- (3,3);
\draw[color=black, thick] (2,4) -- (4,4);
\draw[color=black, thick] (4,4) -- (4,2);

\draw[color=black, thick] (1,1) -- (2,2);
\draw[color=black, thick] (3,1) -- (4,2);
\draw[color=black, thick] (1,3) -- (2,4);
\draw[color=black, thick] (3,3) -- (4,4);

\draw[blue,fill=blue] (1,1) circle (.5ex);
\draw[blue,fill=blue] (3,1) circle (.5ex);
\draw[blue,fill=blue] (2,2) circle (.5ex);
\draw[blue,fill=blue] (1,3) circle (.5ex);

\draw[red,fill=red] (4,4) circle (.5ex);
\draw[red,fill=red] (2,4) circle (.5ex);
\draw[red,fill=red] (4,2) circle (.5ex);
\draw[red,fill=red] (3,3) circle (.5ex);

\node at (1,0.7) {\footnotesize{111}};
\node at (3,0.7) {\footnotesize{211}};
\node at (0.6,3) {\footnotesize{121}};
\node at (4.2,1.75) {\footnotesize{212}};
\node at (2.2,1.75) {\footnotesize{112}};
\node at (3.3,2.75) {\footnotesize{221}};
\node at (1.6,4) {\footnotesize{122}};
\node at (4.4,4) {\footnotesize{222}};

% \node at (1.5,1.9) {\footnotesize{$\varepsilon = 2$}};
% \node at (2.0,0.7) {\footnotesize{$\varepsilon = 1$}};
% \node at (3.0,1.9) {\footnotesize{$\varepsilon = 1.5$}};
% \node at (4,1.3) {\footnotesize{$\varepsilon = 2$}};
\end{tikzpicture}
    \caption{Heterogeneous DP}
    \label{fig: ex graphs a}
\end{subfigure}
\begin{subfigure}[t]{0.30\textwidth}
    \centering
    \begin{tikzpicture}[scale=0.8]

\draw[color=black, thick] (1,1) -- (1,3);
\draw[color=black, thick] (2,2) -- (2,4);
\draw[color=black, line width=0.75mm] (1,1) -- (3,1);
\draw[color=black, line width=0.75mm] (2,2) -- (4,2);

\draw[color=black, thick] (3,1) -- (3,3);
\draw[color=black, line width=0.75mm] (1,3) -- (3,3);
\draw[color=black, line width=0.75mm] (2,4) -- (4,4);
\draw[color=black, thick] (4,4) -- (4,2);

\draw[color=black, thick] (1,1) -- (2,2);
\draw[color=black, thick] (3,1) -- (4,2);
\draw[color=black, thick] (1,3) -- (2,4);
\draw[color=black, thick] (3,3) -- (4,4);

\draw[blue,fill=blue] (1,1) circle (.5ex);
\draw[blue,fill=blue] (3,1) circle (.5ex);
\draw[blue,fill=blue] (2,2) circle (.5ex);
\draw[blue,fill=blue] (1,3) circle (.5ex);

\draw[red,fill=red] (4,4) circle (.5ex);
\draw[red,fill=red] (2,4) circle (.5ex);
\draw[red,fill=red] (4,2) circle (.5ex);
\draw[red,fill=red] (3,3) circle (.5ex);

\node at (1,0.7) {\footnotesize{111}};
\node at (3,0.7) {\footnotesize{211}};
\node at (0.6,3) {\footnotesize{121}};
\node at (4.2,1.75) {\footnotesize{212}};
\node at (2.2,1.75) {\footnotesize{112}};
\node at (3.3,2.75) {\footnotesize{221}};
\node at (1.6,4) {\footnotesize{122}};
\node at (4.4,4) {\footnotesize{222}};

% \node at (1.5,1.9) {\footnotesize{$\varepsilon = 2$}};
% \node at (2.0,0.7) {\footnotesize{$\varepsilon = 1$}};
% \node at (3.0,1.9) {\footnotesize{$\varepsilon = 1.5$}};
% \node at (4,1.3) {\footnotesize{$\varepsilon = 2$}};
\end{tikzpicture}
    \caption{Personalized DP}
    \label{fig: ex graphs b}
\end{subfigure}
\begin{subfigure}[t]{0.30\textwidth}
    \centering
    \begin{tikzpicture}[scale=0.8]

\draw[color=black, thick] (1,1) -- (3,1);
\draw[color=black,  line width=0.75mm] (2,2) -- (4,2);

\draw[color=black, thick] (1,1) -- (2,2);
\draw[color=black, thick] (3,1) -- (4,2);

\draw[color=black,  line width=0.15mm] (1,1) -- (4,2);
\draw[color=black,  line width=0.15mm] (2,2) -- (3,1);

\draw[color=black, thick] (1,1) -- (2,2);
\draw[color=black, thick] (3,1) -- (4,2);

\draw[blue,fill=blue] (1,1) circle (.5ex);
\draw[blue,fill=blue] (3,1) circle (.5ex);
\draw[blue,fill=blue] (2,2) circle (.5ex);
\draw[red,fill=red] (4,2) circle (.5ex);

\node at (1,0.7) {\footnotesize{111}};
\node at (3,0.7) {\footnotesize{211}};
\node at (4.2,1.75) {\footnotesize{212}};
\node at (1.4,2.0) {\footnotesize{112}};

 \node at (1.2,1.5) {\footnotesize{$\varepsilon_1$}};
 \node at (2.0,0.7) {\footnotesize{$\varepsilon_1$}};
 \node at (3.0,2.2) {\footnotesize{$\varepsilon_2$}};
 \node at (3.7,1.3) {\footnotesize{$\varepsilon_1$}};
  \node at (2.5,1.5) {\tiny{$\leq \varepsilon_1+\varepsilon_2$}};
\end{tikzpicture}

% \begin{tikzpicture}[scale=1]

% \draw[color=black, thick] (1,2) -- (3,2);
% \draw[color=black, thick] (2,3) -- (3,2);
% \draw[color=black, thick] (1,2) -- (2,3);

% \draw[blue,fill=blue] (2,3) circle (.5ex);
% \draw[blue,fill=blue] (3,2) circle (.5ex);
% \draw[blue,fill=blue] (1,2) circle (.5ex);
% \draw[white,fill=white] (1,1) circle (.5ex);

%  \node at (0.9,2.5) {\footnotesize{$\varepsilon = 0.1$}};
%  \node at (2.0,1.7) {\footnotesize{$\varepsilon(u,v) \leq 0.2$}};
%  \node at (3.2,2.5) {\footnotesize{$\varepsilon = 0.1$}};
 
%  \node at (0.7,2.0) {\footnotesize{$u$}};
%  \node at (2.0,3.3) {\footnotesize{$w$}};
%  \node at (3.3,2.0) {\footnotesize{$v$}};
 
% \end{tikzpicture}
    \caption{$d$-metric privacy}
    \label{fig: ex graphs b}
\end{subfigure}
\caption{Examples of heterogeneous DP in this paper,  personalized DP \cite{personalized}, and $d$-metric DP \cite{dmetric}. Each dataset $v = (v^1, v^2, v^3)$ contains three votes, $v^i \in \{1,2\}$. The query is the majority vote which is color coded as \texttt{blue} or \texttt{red}. A higher privacy $\varepsilon_2 < \varepsilon_1$ is schematically shown with thicker lines . Imagine the first person's vote needs to better protected, \emph{but only if} it is the deciding vote. In heterogeneous DP, it is possible to assign higher privacy only on edges $(121)-(221)$ and $(112)-(212)$. In personalized DP, all datasets in which the first person's vote is different must be assigned the same privacy parameter. In $d$-metric privacy, privacy conditions between all datasets must be pre-specified subject to the triangular inequality.}
\label{fig: ex graphs}
\end{figure*}

In \cite{dmetric}, the authors define $\varepsilon(u,v)$ \emph{for all} $u,v \in V$ and assume $d(\cdot)$ is a metric function, which satisfies the triangle inequality $\varepsilon(u,v)\leq \varepsilon(u,w) + \varepsilon(w,v)$ for all $u,v, w \in V$. Here, we define $\varepsilon(u,v)$ for neighboring vertices $u \sim v$ only. See the examples in Fig. \ref{fig: ex graphs} that clarify these distinctions.

\section{Main Results}\label{sec:main}

Let $\mathcal{G}$ be a graph with privacy function $\varepsilon(\cdot)$. Let $T$ be a binary-valued query function on $V$ and $p: S \to [0,1]$ be a \emph{partial} function such that $S \subseteq V$ contains the boundary set $\partial _T (\mathcal{G})$. An \emph{extension} of $p$ is a function $\tilde{p}: V \to [0,1]$ such that for all $v \in S$, we have $ \tilde{p}(v) = p(v)$. The main result of this paper is as follows. 
\begin{theorem}\label{theorem:main1}
There exists a polynomial time algorithm which takes a graph $\mathcal{G}$, a privacy function $\varepsilon(\cdot)$, a binary-valued query $T$, and a partial function $p$ as input and outputs as follows. If $p$ cannot be extended to an $\varepsilon(\cdot)$-DP mechanism, it outputs ``no $\varepsilon$-DP extension of $p$ exists". Otherwise, it outputs an extension $\tilde{p}$ of $p$ which is $\varepsilon(\cdot)$-DP and is also the unique optimal with respect to the $\leq_T$ ordering.
\end{theorem}

Before we outline the main steps of the proof of Theorem~\ref{theorem:main1}, we explain some helpful facts. 
\subsection{Initial DP Conditions:}

The notion of the $\varepsilon(\cdot)$-DP on $\mathcal{G}$ is a \emph{local} property in the sense that for a mechanism to be $\varepsilon(\cdot)$-DP, certain local conditions must be satisfied. For binary-valued heterogeneous mechanisms, this is captured in Definition \ref{def:binhetDP}.

\begin{definition}[Binary-valued heterogeneous differential privacy]\label{def:binhetDP}
The function $p:V\to [0,1]$ is called binary-valued heterogeneous $\varepsilon(\cdot)$-differentially private if, for every $u\sim v$, the following conditions are satisfied:
 \begin{align} \label{eq: r1}
	p(u) &\leq e^{\varepsilon(u,v)} p(v), \\
	1-p(u) &\leq e^{\varepsilon(u,v)}(1- p(v)).\label{eq: r2}
	\end{align}
 Furthermore, by the symmetry, the same set of inequalities hold if we interchange $u$ and $v$:
\begin{align} \label{eq: r3}
	p(v) &\leq e^{\varepsilon(u,v)} p(u),\\ 
	1-p(v) &\leq e^{\varepsilon(u,v)}(1- p(u)).\label{eq: r4}
	\end{align}
We refer to \eqref{eq: r1}-\eqref{eq: r4} as the \emph{initial DP conditions}.
\end{definition}
Note that \eqref{eq: r1}-\eqref{eq: r4} can be rearranged to obtain two upper bounds on $p(u)$ and two lower bounds on $p(u)$. In particular, \eqref{eq: r4} can be rewritten as
\begin{align} \label{eq: 0DPW3}
p(u) &\leq \frac{ p(v) -1 + e^{\varepsilon(u,v)} } {e^{\varepsilon(u,v)}}.
 \end{align}
That is, \eqref{eq: r1} and \eqref{eq: 0DPW3} impose two upper bounds on the mechanism $p(u)$. Because of this simple fact and also the fact that neighboring relationship is a symmetric relationship, we may only consider the upper bounds. That is, if all the upper bound inequalities hold for every adjacent pair of vertices, the lower bounds automatically hold. 

\subsection{Strongest Induced DP Condition:} 
Despite the fact that an $\varepsilon(\cdot)$-DP mechanism has a very localized definition as described above, it must also satisfy \emph{less local} conditions as follows. If two vertices $u,v$ are not neighbors, but both satisfy local $\varepsilon(\cdot)$-DP conditions with a third mutual neighboring vertex $u_0$, then a weaker set of inequalities must hold for the value of the mechanism at $u$ and $v$. As we take different  $(u,v)$-paths, still certain inequalities, depending on that path, on the value of the mechanism at the two end vertices must be satisfied. 

The upper bounds a vertex imposes on another vertex via some path are called \emph{induced DP conditions}. However, as the path is incremented, the induced conditions become weaker.

In fact, for every pair of nodes $u,v \in V$, any $\varepsilon(\cdot)$-DP mechanism on $\mathcal{G}$, and any path $\rho \in \mathcal{P}(u,v)$, the vertex $u$ imposes an upper bound on $p(v)$, which depends on both $p(u)$ and $\rho$. We denote this upper bound by $U_{\rho , p(u)}(v)$. Thus, from now on, every time we talk about the best bound $u$ imposes on $p(v)$, we mean the smallest upper bound on $p(v)$ among all $\rho \in \mathcal{P}(u,v)$, i.e., $\min_{\rho\in \mathcal{P}(u,v)} U_{\rho , p(u)}(v)$. We call this smallest bound \emph{the strongest induced condition of $u$ on $p(v)$.} The following lemma summarizes these facts.
%, which 
 
\begin{lemma} \label{lem: increasing} Let $\rho$ be a $(u,v)$-path and $v'$ be a middle vertex of this path. Let $\rho'$ and $\rho''$ be $(u,v')$ and $(v',v)$ subpaths on $\rho$. Then, for the upper bound $U$ function we obtain the expression $U_{\rho, p(u)}(v) = U_{\rho'', U_{\rho', p(u)}(v')}(v)$.
Also, for every $\varepsilon(\cdot)$-DP mechanism $p$, the upper bounds $u$ imposes on $p(v')$ are smaller than the ones it imposes on $p(v)$.
\end{lemma}

The above lemma follows from Theorem \ref{theo:path} that will appear later in the paper.

\subsection{Finding the Strongest Induced DP Condition}

For a single path graph, finding the strongest induced DP condition can be solved efficiently. The closed-form expression of the optimal induced DP conditions for a path graph is derived in Theorem \ref{theo:path}.

However, finding the strongest induced condition is challenging in a general graph, since the number of the paths in the graph can be exponentially large on the number of vertices. However, in Algorithm \ref{algo: acomplete}, we show that this task can be accomplished in polynomial time.  In the next section, we describe this algorithm, prove its correctness, and analyze its time complexity.

We now explain the proof of Theorem \ref{theorem:main1} and the construction of Algorithm \ref{algo: acomplete}.

\begin{itemize}
\item{\textbf{Checking the extendibility:}} The idea is very simple. We find a necessary condition that any extendable function must satisfy to be  $\varepsilon(\cdot)$-DP and we check it in polynomial time. If it is not satisfied, the algorithm declares that no $\varepsilon(\cdot)$-DP extension exists.
%\varepsilon(\cdot)$

Actually, the necessary conditions are the induced DP conditions over the pairs $u,v\in S$. In fact, all the induced conditions are necessary, but since we only have access to partial $p$, over the set $S$, we only consider those ones. Note that it is enough to only check the strongest induced conditions. The number of the ordered pairs in $S$ is $|S|\cdot(|S|-1)$ and hence, polynomial. Also checking if the strongest condition is satisfied over any pair in $S$ can be done using Algorithm \ref{algo: acomplete}.

\item{\textbf{Finding an extension:}} Note that if the necessary condition is satisfied in the previous part, then  Algorithm \ref{algo: acomplete} outputs some extension function of $p$. If we can prove that the output in this case is indeed an $\varepsilon(\cdot)$-DP mechanism, firstly it shows that the checking extendibility is indeed sufficient, and more importantly, the output satisfies $\varepsilon(\cdot)$-DP. 

The way the algorithm extends the function $p$ over the entire $V$ is as follows. For a vertex $v$ outside $S$, and based on the value of $T(v)$ it does the following. When $T(v)=1$, the algorithm searches over all the vertices $u\in S$ and finds the one whose strongest induced condition on the vertex $v$ is the smallest. In other words, among all the paths of the form $\rho \in \mathcal{P}(u,v)$ in which $u\in S$, it takes the one with minimum possible $U_{\rho , p(u)}(v)$. For any fixed $u$, this can be done by using Algorithm \ref{algo: subroutine}. Since $|S|$ has also polynomial size, the entire task can be done in polynomial time.

When $T(v)=2$, the symmetry in the problem can be used to perform exactly the same steps except that we call Algorithm \ref{algo: acomplete} with the input $1-p$ instead of $p$. In Section \ref{TSaP.B}, we prove that the resulting extension of $p$ is $\varepsilon(\cdot)$-DP. 

%change: resulting $p$ --> 

\item{\textbf{Optimality with respect to $\leq_T$:}} This part is a direct consequence of the previous item. The reason is that for every vertex $v\in V \setminus S$, depending on whether $T(v)=1$ or $2$, we assigned the minimum upper bound (respectively, maximum lower bound) that is necessary for any $\varepsilon(\cdot)$-DP mechanism. In other words, if we increase the value of $p(v)$ (for the case $T(v)=1$) by any positive value, then one necessary condition fails and therefore the function cannot be $\varepsilon(\cdot)$-DP. 
\end{itemize}

\section{Technical Statements and Proofs}\label{sec:technical}
\subsection{The Path Graph}
Let $\rho = v_0, \cdots, v_n$ be a path graph of length $n$ with the mechanism specified only at the head node $v_0$, i.e., $p(v_0) = \Pr[\mathcal{M}(v_0) = 1] =\alpha$ is given. The query value at $v_0$ can be either $T(v_0) = 1$ or $T(v_0) = 2$, but it is assumed $T(v_i) = 1$ across every other node on the path. This ensures that extending $p(v_0)$ in an optimal manner across the rest of the path is equivalent to maximizing $p(v_i)$, for $1 \leq i \leq n$. Since we are dealing with a path graph only and for simplicity of notation, we use the shorthand $\varepsilon_i \coloneqq\varepsilon(v_i,v_{i+1})$ here. Therefore, \eqref{eq: r1} and \eqref{eq: 0DPW3} are re-written for $i = 0, \cdots, n-1$ as 
\begin{align}\label{eq:  d=0, ex+d}
p(v_{i+1}) &\leq  e^{\varepsilon_i} p(v_{i}),\\
p(v_{i+1}) &\leq \frac{ p(v_{i}) -1 + e^{\varepsilon_i}}{e^{\varepsilon_i}} \label{eq:  d=0, e(1-a)+d}.
\end{align}

Among the two upper bounds \eqref{eq:  d=0, ex+d} and \eqref{eq:  d=0, e(1-a)+d} on $p(v_{i+1})$, the smallest one prevails at dictating the optimal mechanism: $$p(v_{i+1}) \leq \min \left\{  e^{\varepsilon_i} p(v_{i}), \frac{ p(v_{i}) -1 + e^{\varepsilon_i}}{e^{\varepsilon_i}}\right\}.$$ We first present a simple lemma that characterizes which upper bound among \eqref{eq:  d=0, ex+d} and \eqref{eq:  d=0, e(1-a)+d} is the tightest. 

\begin{lemma}\label{lem: iff}
A necessary and sufficient condition for 
\[\min \left\{  e^{\varepsilon_i} p(v_{i}), \frac{ p(v_{i}) -1 + e^{\varepsilon_i}}{e^{\varepsilon_i}}\right\} = e^{\varepsilon_i} p(v_{i})
\]
is that $p(v_i) \leq  (e^{\varepsilon_i}+ 1)^{-1}$.
\end{lemma}

\begin{IEEEproof}
\begin{align*}
e^{\varepsilon_i} p(v_i) \leq  \frac{ p(v_i) -1 + e^{\varepsilon_i}}{e^{\varepsilon_i}} &\iff\\
e^{2\varepsilon_i} p(v_i)   \leq p(v_i) -1 + e^{\varepsilon_i}  &\iff \\
e^{2\varepsilon_i} p(v_i) -p(v_i)  \leq  e^{\varepsilon_i} -1  &\iff \\
(e^{2\varepsilon_i}-1) p(v_i)  \leq  e^{\varepsilon_i} -1  &\iff
p(v_i)  \leq  \frac{1}{e^{\varepsilon_i}+1}
\end{align*}
where the last step follows from $e^{\varepsilon_i} -1 \geq 0$. Note, that if $e^{\varepsilon_i}-1 = 0$ then, the inequality is trivial.\footnote{Note that Lemma \ref{lem: iff} is true for the general graph and general $\varepsilon(u,v)$.}
\end{IEEEproof}

The optimal binary-valued heterogeneous differentially private mechanism $p^*$ with initial condition $p^*(v_0) = \alpha$, follows from Lemma \ref{lem: iff} and induction on $i$, and is given by\footnote{We check if \eqref{eq: optimalp} gives a value greater than 1 and if so, set $ p^*(v_{i+1}) = 1$.} %i.}
 \begin{align}\label{eq: optimalp}
 p^*(v_{i+1}) = \min \left\{  e^{\varepsilon_i} p^*(v_{i}), \frac{ p^*(v_{i}) -1 + e^{\varepsilon_i}}{e^{\varepsilon_i}}\right\}.
 \end{align}
 
\begin{theorem}\label{theo: pathupperboundary}
 The function $p^*$ given in \eqref{eq: optimalp} is the unique optimal $\varepsilon(\cdot)$-DP mechanism on the path graph $\rho$.
\end{theorem}
\begin{IEEEproof}
    Assume that $p^*$ is not optimal. Let $ p':V\to [0,1]$ be another binary-valued heterogeneous $\varepsilon(\cdot)$-differentially private function and let $k \in \mathbb{N}$ be the smallest number such that $p^*(v_k) < p'(v_k)$.

    As $k$ is the smallest number which satisfies the statement above, we obtain $ p'(v_{k-1}) \leq p^*(v_{k-1}) $. This means we have
    \begin{align}\label{eq:intermediate1}
    e^{\varepsilon_{k-1}} p'(v_{k-1})  &\leq  e^{\varepsilon_{k-1}} p^*(v_{k-1}),\\
    \frac{ p'(v_{k-1}) -1 + e^{\varepsilon_{k-1}}}{e^{\varepsilon_{k-1}}} &\leq \frac{ p^*(v_{k-1}) -1 + e^{\varepsilon_{k-1}}}{e^{\varepsilon_{k-1}}}.\label{eq:intermediate2} 
    \end{align}

    Since $p'$ is an  $\varepsilon(\cdot)$-differentially private function, it satisfies \eqref{eq:  d=0, ex+d} and \eqref{eq:  d=0, e(1-a)+d}. Combining this with \eqref{eq:intermediate1} and \eqref{eq:intermediate2} we obtain
\begin{align}
  p'(v_{k})  &\leq 
    \min \left\{e^{\varepsilon_{k-1}} p'(v_{k-1})  , \frac{ p'(v_{k-1}) -1 + e^{\varepsilon_i}}{e^{\varepsilon_{k-1}}} \right\}\\
    &\leq 
    \min \left\{e^{\varepsilon_i} p^*(v_{k-1}),\frac{ p^*(v_{k-1}) -1 + e^{\varepsilon_{k-1}}}{e^{\varepsilon_{k-1}}}\right\} \\&= p^*(v_{k}),    
    \end{align}
    where the last equality is due to the construction of $p^*$ in Theorem \ref{theo: pathupperboundary}. This contradicts the first assumption.
    The proof of uniqueness is straightforward.
\end{IEEEproof}

We now show how to find the optimal $\varepsilon(\cdot)$-DP mechanism for the path graph.
%$\varepsilon(\cdot)$-DP

\begin{theorem}\label{theo:path}
Let the path graph $\rho = v_0, \cdots, v_n$ and the mechanism at its head vertex $p(v_0) = \alpha$ be given. Assume that $T(v_i) = 1$ and $\varepsilon_i > 0$ is fixed, for $1\leq i\leq n$. Then, the optimal binary-valued heterogeneous differentially private mechanism $p^*$ is given by
\begin{align*}
  p^*(v_i) \coloneqq  
  \begin{cases}
    e^{\varepsilon_{i-1}+\cdots + \varepsilon_{0}} \alpha, &    i \leq \tau,  \\ 
    e^{-\varepsilon_{i-1}-\dots - \varepsilon_{\tau+1} - \varepsilon_{\tau} +\varepsilon_{\tau-1}\dots + \varepsilon_{0}} \alpha &  i > \tau. \\
       \quad + 1 - e^{-\varepsilon_{i-1}-\dots - \varepsilon_{\tau}},   
    \end{cases}
\end{align*}
where
\begin{align}\label{eq:tau0}
 \tau &= \argmin_{i \in [n]}\left\{\frac{1}{\alpha}\leq e^{\varepsilon_{i-1}+\varepsilon_{i-2} +\dots + \varepsilon_{0}}(e^{\varepsilon_{i}}+1)\right\}.
 \end{align}
\end{theorem}
\begin{IEEEproof}
First assume that there exists some $k \in [n]$ such that for all $1 \leq  i < k$, we have $\min \{  e^{\varepsilon_i} p(v_{i}), \frac{ p(v_{i}) -1 + e^{\varepsilon_i}}{e^{\varepsilon_i}}\} = e^{\varepsilon_i} p(v_{i})$. That is, \eqref{eq:  d=0, ex+d} is the tightest upper bound on $p(v_{i+1})$. We will soon find the largest $k$ for which this can happen. Iterating over $i = k, k-1, \cdots, 1$, we will construct the mechanism $p^*$ through induction
\begin{align}\label{eq:last1}
    \mathllap p^*(v_{k}) & =  e^{\varepsilon_{k-1}} p^*(v_{k-1})\\
      & = e^{\varepsilon_{k-1}}\big{(}e^{\varepsilon_{k-2}} p^*(v_{k-2})\big{)}\\
      & \cdots\\
      & = e^{\varepsilon_{k-1}}\bigg{(}
      e^{\varepsilon_{k-2}}
      \Big{(} \cdots
      \big{(}e^{\varepsilon_{0}} p(v_{0}) \big{)}
      \cdots
      \Big{)} \bigg{)}\\
      &= e^{\varepsilon_{k-1}+\varepsilon_{k-2} +\dots + \varepsilon_{0}} \alpha.\label{eq:lastn}
  \end{align}
We now want to find the smallest index on the path for which \eqref{eq:  d=0, ex+d} is not tight. This is, we want to find the last index for which the iterations \eqref{eq:last1}-\eqref{eq:lastn} hold. Let this parameter to be $\tau$. Therefore, on the one hand, $\tau$ satisfies
\begin{align}\label{eq:tau1}
p^*(v_{\tau}) &  = e^{\varepsilon_{\tau-1}+\varepsilon_{\tau-2} +\dots + \varepsilon_{0}}\alpha.
\end{align}
On the other hand, by the definition of $\tau$, for $i = \tau + 1$, \eqref{eq:  d=0, e(1-a)+d} will give the tightest upper bound on $p(v_{\tau+1})$. That is, $$\min \left\{  e^{\varepsilon_\tau} p^*(v_{\tau}), \frac{ p^*(v_{\tau}) -1 + e^{\varepsilon_\tau}}{e^{\varepsilon_\tau}}\right\} = \frac{ p^*(v_{\tau}) -1 + e^{\varepsilon_\tau}}{e^{\varepsilon_\tau}}.$$ Therefore, from Lemma \ref{lem: iff}, we must have
\begin{align}\label{eq:tau2}
p^*(v_{\tau}) \geq \frac{1}{ e^{\varepsilon_{\tau}} + 1}. 
  \end{align}
 Combining \eqref{eq:tau2} and  \eqref{eq:tau1} and taking the minimum over all $i\in [n]$ gives \eqref{eq:tau0}. 
 
We need to verify that the upper bounds \eqref{eq:  d=0, ex+d} and  \eqref{eq:  d=0, e(1-a)+d} do not ``toggle" or ``alternate" in providing the tightest bound on $p^*(v_i)$ for $i > \tau$. Referring to Lemma \ref{lem: iff}, this is equivalent to verifying that for every $\tau \leq i \leq n$ we will have
\[p^*(v_{i}) \geq \frac{1}{ e^{\varepsilon_{i}} + 1}.\]
For $i = \tau$ this holds by definition. For $\tau < i \leq n$, this can be proved via contradiction. Assume there exists $i> \tau$ such that the following statements are satisfied.
\begin{align}\label{eq:firsttau}
p^*(v_{i}) &\geq \frac{1}{ e^{\varepsilon_{i}} + 1},\\
p^*(v_{i+1}) &< \frac{1}{ e^{\varepsilon_{i+1}} + 1}.\label{eq:secondtau}
\end{align}
Since \eqref{eq:firsttau} is satisfied, from Lemma \ref{lem: iff}, we must have:
\begin{align}
    p^*(v_{i+1})&= \min \left\{  e^{\varepsilon_i} p^*(v_{i}), \frac{ p^*(v_{i}) -1 + e^{\varepsilon_i}}{e^{\varepsilon_i}}\right\} \\&= \frac{ p^*(v_{i}) -1 + e^{\varepsilon_{i}}}{e^{\varepsilon_{i}}}.
\end{align}
Therefore, substituting $p^*(v_{i+1})$ with the above equation in \eqref{eq:secondtau} leads us to:
\begin{align} \label{eq: <1}
p^*(v_{i+1}) = \frac{ p^*(v_{i}) -1 + e^{\varepsilon_{i}}}{e^{\varepsilon_{i}}}  <\frac{1}{ e^{\varepsilon_{i+1}} + 1}.
\end{align}
Using the bound in \eqref{eq:firsttau} on $p^*(v_i)$ in the above gives
\[
\frac{ \frac{1}{ e^{\varepsilon_{i}} + 1} -1 + e^{\varepsilon_{i}}}{e^{\varepsilon_{i}}} < \frac{1}{ e^{\varepsilon_{i+1}} + 1},
\]
which is equivalent to:
\begin{align}
\frac{ \frac{1+ e^{2\varepsilon_{i}} - 1}{ e^{\varepsilon_{i}} + 1}} {e^{\varepsilon_{i}}}
&< \frac{1}{ e^{\varepsilon_{i+1}} + 1}\iff \\
\frac{ e^{\varepsilon_{i}}} { e^{\varepsilon_{i}} + 1}
&< \frac{1}{ e^{\varepsilon_{i+1}} + 1} \iff 
 e^{\varepsilon_{i}} e^{\varepsilon_{i+1}}  <  1.
 \end{align}
The last statement is a contradiction as $ 0< \varepsilon_{i} + \varepsilon_{i+1}$. 

Having proved that the optimal mechanism has at most two regimes, as  determined by a single $\tau$, the last step is to provide a closed-form expression for the iterations $\tau < i \leq n $. Starting with $i = \tau+1$, we will have
\begin{align}
p^*(v_{\tau+1}) &=  \frac{ p^*(v_{\tau}) -1 + e^{\varepsilon_\tau}}{e^{\varepsilon_\tau}}\\&=\frac{e^{\varepsilon_{\tau-1}+\varepsilon_{\tau-2} +\dots + \varepsilon_{0}}\alpha-1+e^{\varepsilon_\tau}}{e^{\varepsilon_\tau}}\\&=e^{-\varepsilon_\tau}e^{\varepsilon_{\tau-1}+\varepsilon_{\tau-2} +\dots + \varepsilon_{0}}\alpha-e^{-\varepsilon_\tau}+1.
\end{align}
For $i = \tau +2$, we will get 
\begin{align*}
p^*(v_{\tau+2}) &=  \frac{ p^*(v_{\tau+1}) -1 + e^{\varepsilon_{\tau+1}}}{e^{\varepsilon_{\tau+1}}}\\&=\frac{ e^{-\varepsilon_\tau+\varepsilon_{\tau-1}+\varepsilon_{\tau-2} +\dots + \varepsilon_{0}}\alpha-e^{-\varepsilon_\tau}+ e^{\varepsilon_{\tau+1}}}{e^{\varepsilon_{\tau+1}}}\\&=
e^{-\varepsilon_{\tau+1}-\varepsilon_\tau+\varepsilon_{\tau-1}+\varepsilon_{\tau-2} +\dots + \varepsilon_{0}}\alpha-e^{-\varepsilon_{\tau+1}-\varepsilon_\tau}+1.
\end{align*}
continuing this for $i > \tau + 2$ completes the proof.
\end{IEEEproof}

We recover the result for the homogeneous case \cite{RafnoDP2colorPaper}.

\begin{corollary}
Let $\varepsilon > 0$ and set $\varepsilon_i = \varepsilon$ for $i \in [n]$. Theorem~\ref{theo:path} recovers the results in \cite{RafnoDP2colorPaper} for $\delta = 0$.
\end{corollary}
\begin{IEEEproof}
 For $\varepsilon_i = \varepsilon$ for $i \in [n]$, the value of $\tau$ from \eqref{eq:tau0} is
\begin{align}\label{eq:tauspecial}
 \tau_1 &= \left\lceil\frac{1}{\varepsilon}\log\left(\frac{1}{\alpha(1+e^{\epsilon})}\right)\right\rceil.
 \end{align} 
 Under $\varepsilon_i = \varepsilon$ for $i \in [n]$, $p^*$ given in Theorem \ref{theo:path} is simplified to
 \begin{align*}
  p^*(v_i) \coloneqq  
  \begin{cases}
    e^{i\varepsilon} \alpha, &  i \leq \tau_1,  \\ 
    1-e^{(i-\tau_1)\varepsilon}+e^{(-i+2\tau_1)\varepsilon}\alpha, &  i > \tau_1.
    \end{cases}
\end{align*}
Note the results in \cite{RafnoDP2colorPaper} were in terms of ``the probability of being red": $R_i = (1-p_i) = \Pr[\mathcal{M}(v_i) = 2]$. Also, the head vertex in \cite{RafnoDP2colorPaper} started at $i = n_B$ instead of $i=0$ here, which is adopted for easier notation in this paper. With appropriate index conversion, it can be verified that we recover the results in Theorem 10 in \cite{RafnoDP2colorPaper} for $\delta = 0$.
\end{IEEEproof}

\subsection{The General Case}\label{TSaP.B}

In this section, we generalize the results of the previous section. We assume for a given general graph $\mathcal{G}$ and heterogeneous privacy budget $\varepsilon(\cdot)$ over $E$, the mechanism is specified a priori over a subset of vertices $S\subseteq V$, such that $\partial _T (\mathcal{G})\subseteq S$, i.e, $p(u) = \alpha_u$, for every $u\in S$, is given with no additional assumptions on $\alpha_u$. Our goal is to extend $p$ for all other vertices whose mechanism is to be specified in an optimal and computationally efficient manner. The following definition will come handy.

\begin{definition}[The path upper bound function]\label{def: The path upper bound function}
For every vertex $v \in \mathcal{G}$ and $(u,v)$-path $\rho \in \mathcal{P}(u,v)$ such that its head node $u\in S$, we define $U_{\rho, \alpha_u} (v)$ to be the upper bound on the value of $p(v)$ imposed by $p(u) = \alpha_u$.
\end{definition}

We prove that the following optimization problem can be solved in polynomial time for every vertex $u$ with a fixed value $\alpha \in [0,1]$.
\[\minimize_{\rho \in \mathcal{P}(u,v)} \quad U_{\rho, \alpha}(v)\]

To this end, we propose the polynomial Algorithm \ref{algo: subroutine} which takes $u,\alpha$ and $ \mathcal{G}$ as the input and outputs $A_{u, \alpha}(v)$ for every vertex $v \in V$ such that $A_{u, \alpha}(v) = \min _{\rho \in \mathcal{P}(u,v)} U_{\rho, \alpha}(v).$

\begin{theorem}\label{theo: aoptimal}
If $A_{u, \alpha}(v)$  is the output of the Algorithm \ref{algo: subroutine} then, 
\[A_{u, \alpha}(v) = \min _{\rho \in \mathcal{P}(u,v)} U_{\rho, \alpha}(v).\]
\end{theorem}

\begin{IEEEproof}
Let $v_1, \dots , v_{n-1}$ be the vertices of graph in the order Algorithm \ref{algo: subroutine} selects.  
By contradiction, let $k$ be the smallest index such that $A_{u, \alpha}(v_k) \not =  \min _{\rho \in \mathcal{P}(u,v_k)} U_{\rho, \alpha}(v).$
First, assume that $A_{u, \alpha}(v_k)  < \min _{\rho \in \mathcal{P}(u,v_k)} U_{\rho, \alpha}(v_k).$ From Algorithm \ref{algo: subroutine} (line 7), we have $A_{u, \alpha}(v_k) = \min_{v \in N(S_k)} \alpha^*(v)$.
Define $u_k$ as:
\[
\begin{aligned}
u_k := \argmin_{v' \in N^*(v_k)} U_{(v',v_k),A_{u, \alpha}(v')}(v_k)\\
\Rightarrow A_{u, \alpha}(v_k) = U_{(u_k,v_k),A_{u, \alpha}(u_k)}(v_k).
\end{aligned}
\]
Then, by the choice of $k$, $u_k $ belongs to $S_k$ and we have:
\[A_{u, \alpha}(u_k) = \min _{\rho \in \mathcal{P}(u,u_k)} U_{\rho, \alpha}(u_k) = U_{\rho', \alpha}(u_k).\]
Then, following from Lemma \ref{lem: increasing} we have:
\[
\begin{aligned}
A_{u, \alpha}(v_k) & = U_{(u_k,v_k),A_{u, \alpha}(u_k)}(v_k) &\\
& = U_{(u_k,v_k),U_{\rho', \alpha}(u_k)}(v_k)& = U_{\rho' v_k,\alpha}(v_k).
\end{aligned}
\]
Thus, $ A_{u, \alpha}(v_k)$ ($=U_{\rho' v_k,\alpha}(v_k)$) cannot be less than $\min _{\rho \in \mathcal{P}(u,v)} U_{\rho, \alpha}(v_k)$. 
\\It remains to consider the case: $A_{u, \alpha}(v_k)  > \min _{\rho:\rho \in \mathcal{P}(u,v_k)} U_{\rho, \alpha}(v_k).$ Let $\rho$ be the $(u,v_k)$-path that imposes the strongest induced condition on $v_k$. First observe that $\rho$ connects a vertex inside $S_k$ (i.e., $u$) to a vertex outside (i.e., $v_k$). Following from Lemma \ref{lem: increasing}, if we traverse this path from head to tail, the first time we leave $S_k$ must be the last step. Since, otherwise we would not have selected $v_k$.
\\Let $w$ be the vertex on $\rho$ before we reach $v_k$. Hence, $w $ is also appeared in $N^*(v_k)$. Therefore, by the choice of $\alpha^*(v_k)$, we have $A_{u, \alpha}(v_k)=\alpha^*(v_k) \leq U_{(w,v_k),A_{u,\alpha}(w)}(v_k)$. The equality is because of the choice of $v_k$ and the inequality is because of the definition of $\alpha^*$ and the fact that $U_{(w,v_k),A_{u,\alpha}(w)}$ is one of the terms in minimizing the problem which defines $\alpha^*(v_k)$.
%minimizing the

\end{IEEEproof}

Let $p: S \to [0,1]$ be a partial function. We want to know under what condition $p$ can be extended to an $\varepsilon$-DP function. To this end, first we define the notion of \say{compatible function} and then we show that the necessary and sufficient condition for $p$ to be extendable to a $\varepsilon$-DP function is the compatibility condition. We also prove that testing compatibility can be done in polynomial time.

\begin{definition}
A partial function $p:S \to [0,1]$ is called compatible if, for every vertices $u,v \in S$, it follows that $p(v) \leq \min _{\rho \in \mathcal{P}(u,v)} U_{\rho, \alpha}(u)$.
\end{definition}

% Now we give the technical proof of Theorem \ref{theorem:main1}
% fixed a few $\varepsilon(\cdot)$
Now we give the technical proof of Theorem \ref{theorem:main1}.
\begin{algorithm}[t] 
    \caption{Construction of the $A$ function}\label{algo: subroutine}
    \SetAlgoLined
    \SetAlgoNoLine
        \SetAlgoLined
    \textbf{Input:} Graph $\mathcal{G} $, $u \in V$, $\alpha_{u} = p(u)$.
    
    \textbf{Output: Function $A_{u, \alpha}: V \to [0,1]$}
   
    \nl $i \gets 1$
    
    \nl $S_1 \gets \{u\}$ 

    \nl $A_{u,\alpha}(u) \gets \alpha_{u} $

    \While{$|S_i|< |V| $}{
    \For{ $v \in N(S_i)$}{
    
    \nl $N^*(v) \gets N(v) \cap S_i $
    
    \nl $\alpha^*(v) \gets \min_{v' \in N^*(v)} U_{(v',v),A_{u,\alpha}(v')}(v)$ 
    }
    
    \nl $v_i \gets \argmin_{v \in N(S_i)} \alpha^*(v)$

    \nl $A_{u, \alpha}(v_i) \gets \min_{v \in N(S_i)} \alpha^*(v)$ 

    \nl $S_{i+1} \gets S_i \cup \{v_i\}$
    
    \nl $i \gets i+1$
    }
\end{algorithm}

\begin{algorithm}[t] 
    \caption{Construction of the extension of a mechanism} \label{algo: acomplete}
    \SetAlgoLined
    \textbf{Input:}  Graph $\mathcal{G} $, subset $S \subseteq V$ which $\partial _T (\mathcal{G}) \subseteq S$, partial function $p: S \to [0,1]$.
    
    \textbf{Output: Function $\tilde{p}: V \to [0,1]$}

    \nl $S_1 \gets S$
%    $\varepsilon(\cdot)$-DP
    \If{$p$ is not compatible}{
    \nl \Return No $\varepsilon(\cdot)$-DP extension of $p$ exists.
    }
    \For{$v \in S $ }{
    \nl $\tilde{p}(v) \gets p(v) $
    }
    
    \For{ $v \in V \setminus S $}{
    \If{$T(v) = 1$}{
    \nl $\tilde{p}(v) \gets \min_{u \in S} A_{u, \tilde{p}(u)}(v)$
    }
    \If{$T(v) = 2$}{
    \nl $\tilde{p}(v) \gets \min_{u \in S} A_{u, 1- \tilde{p}(u)}(v)$
    }
    }
    
\end{algorithm}

\begin{IEEEproof}[Proof of Theorem \ref{theorem:main1}]
Since $\tilde{p}$ is an extension function of $p$, for the sake of simplicity, in the rest of the proof, we denote $\tilde{p}$ by $p$.
In Section \ref{HDP}, we argued that to complete the proof of Theorem \ref{theorem:main1}, we must show that if Algorithm \ref{algo: acomplete} outputs a function $p$, then it is an $\varepsilon(\cdot)$-DP mechanism.
By the definition, to show that a function is $\varepsilon(\cdot)$-DP, we must verify \eqref{eq: r1} and \eqref{eq: r2} hold for every edge $(u, v)$ in $E$. We consider three cases.
\begin{itemize}
\item{\textbf{Case 1: ($u,v \in S$)}}
%therefore
In this case, since the algorithm has passed the compatibility test, the value of $p$ at $u,v$ are consistent; that is, $p(v)$ is no more than the strongest condition $p(u)$ imposes on it via all possible $(u, v)$-paths. In particular, the one imposed by the edge $(u,v)$ is also guaranteed. Similarly, the condition $p(v)$ imposes on $p(u)$ must be satisfied. Thus, the initial conditions on the edge $(u,v)$ are satisfied.

\item{\textbf{Case 2: ($u\in S, v\notin S$)}}
First, because of the symmetry and without loss of generality, let us assume that $T(v)=1$. In this case, similar to the earlier case, we can observe that the condition $p(u)$ imposes on $p(v)$ is satisfied. We just have to show that the converse is also true.
Assuming the opposite we have $U_{(v,u), p(v) } (u) < p(u)$. 
Since $p(v)$ is assigned by the algorithm as the least upper bounds imposed by all the vertices in $S$ via all the possible paths connecting them to $v$, let $\omega$ and $\rho_1$ be the vertex and the path that give the least upper bound to $p(v)$. Then, following from Lemma \ref{lem: increasing} we have:
\begin{align}\label{eq: T5,e2}
\begin{aligned}
U_{\rho_1u,  p(\omega) } (u) & = U_{(v,u),  [U_{\rho_1,  p(\omega) } (v) ]} (u) \\& = U_{(v,u),  p(v)} (u) .
\end{aligned}
\end{align}

Following from the compatibility condition, as $\omega, u \in S$:
\begin{align}\label{eq: T5,e1}
p(u) \leq U_{\rho_1u,  p(\omega) } (u).
\end{align}

Finally, \eqref{eq: T5,e2} and \eqref{eq: T5,e1}  lead us to the following contradiction: $p(u)  \leq U_{\rho_1u,  p(\omega) } (u)  = U_{(v,u),  p(v)} (u)  < p(u) $.

% \eqref{eq: T5,e2} and \eqref{eq: T5,e1} 
%, which is a contradiction

\item{\textbf{Case 3: ($u,v \notin S$)}} In this case, we first observe that $T(u)=T(v)$, since if it is not the case, then $u$ and $v$ are boundary vertices and therefore by the assumption, $u,v\in S$, which is a contradiction. Now, without loss of generality, let us assume that $T(u)=T(v)=1$ and $U_{(v,u), p(v) } (u) < p(u)$. Then, from the same setting of the previous case we obtain the following equation: $ U_{\rho_1u, p(\omega)}(u)  = U_{(v,u), p(v) }  (u) < p(u) $.
Also, Algorithm \ref{algo: acomplete} (line 13) implies that $p(u) = \min_{\omega' \in S} A_{\omega', p(\omega')}(u)$ and, Theorem \ref{theo: aoptimal} leads us to the following: \[ p(u) = \min_{\omega' \in S} A_{\omega', p(\omega')}(u) = \min_{\omega' \in S, \rho \in \mathcal{P}(\omega',u)} U_{\rho, p(\omega')}(u),\]
which contradicts $U_{\rho_1u, p(\omega)}(u)< p(u)$.
\end{itemize}
The proof of uniqueness is straightforward.
The last part in Theorem \ref{theorem:main1} is regarding the time complexity of Algorithm \ref{algo: acomplete}. This is discussed in the next part.
%also
\end{IEEEproof}

Now, we analyze the running time of Algorithms. In Algorithm \ref{algo: subroutine} at each iteration, every edge which has exactly one endpoint in $S$ is considered, and the one with the best (minimum) upper bound on the other end is selected. This will take at most $\mathcal{O}(|E|)$ time. Thus, Algorithm \ref{algo: subroutine} runs in $\mathcal{O}(|E|\cdot|V|)$ time. In Algorithm \ref{algo: acomplete},  compatibility test takes $\mathcal{O}(|V|^2)$ calls of  Algorithm \ref{algo: subroutine}. If $p$ is compatible, then for every vertex $v\in V\setminus S$ we check the best bound. For that, we will call Algorithm \ref{algo: subroutine} $\mathcal{O}(|S| \cdot |V\setminus S| = \mathcal{O}(|V|^2) $ many times. In conclusion, we have the following theorem.

\begin{theorem}
Algorithm \ref{algo: subroutine} and \ref{algo: acomplete} run in $\mathcal{O}(|E| |V|)$ and $\mathcal{O}(|V|^3|E|)$ respectively.
\end{theorem}

%\bibliographystyle{IEEEtran}
%\bibliography{ref}

\begin{thebibliography}{10}
\providecommand{\url}[1]{#1}
\csname url@samestyle\endcsname
\providecommand{\newblock}{\relax}
\providecommand{\bibinfo}[2]{#2}
\providecommand{\BIBentrySTDinterwordspacing}{\spaceskip=0pt\relax}
\providecommand{\BIBentryALTinterwordstretchfactor}{4}
\providecommand{\BIBentryALTinterwordspacing}{\spaceskip=\fontdimen2\font plus
\BIBentryALTinterwordstretchfactor\fontdimen3\font minus
  \fontdimen4\font\relax}
\providecommand{\BIBforeignlanguage}[2]{{%
\expandafter\ifx\csname l@#1\endcsname\relax
\typeout{** WARNING: IEEEtran.bst: No hyphenation pattern has been}%
\typeout{** loaded for the language `#1'. Using the pattern for}%
\typeout{** the default language instead.}%
\else
\language=\csname l@#1\endcsname
\fi
#2}}
\providecommand{\BIBdecl}{\relax}
\BIBdecl

\bibitem{dwork2006calibrating}
C.~Dwork, F.~McSherry, K.~Nissim, and A.~Smith, ``Calibrating noise to
  sensitivity in private data analysis,'' in \emph{Proc. Theory Cryptography
  Conf.}, New York, NY, Mar. 2006, pp. 265--284.

\bibitem{survey_2017}
T.~Zhu, G.~Li, W.~Zhou, and P.~S. Yu, ``Differentially private data publishing
  and analysis: A survey,'' \emph{IEEE Transactions on Knowledge and Data
  Engineering}, vol.~29, no.~8, pp. 1619--1638, 2017.

\bibitem{Censusissues}
S.~L. Garfinkel, J.~M. Abowd, and S.~Powazek, ``Issues encountered deploying
  differential privacy,'' in \emph{Proceedings of the 2018 Workshop on Privacy
  in the Electronic Society}, 2018, pp. 133--137.

\bibitem{USCBadopts}
J.~M. Abowd, ``The {US Census Bureau} adopts differential privacy,'' in
  \emph{Proceedings of the 24th ACM SIGKDD International Conference on
  Knowledge Discovery \& Data Mining}, 2018, pp. 2867--2867.

\bibitem{epsilonvote}
N.~Kohli and P.~Laskowski, ``Epsilon voting: Mechanism design for parameter
  selection in differential privacy,'' in \emph{2018 IEEE Symposium on
  Privacy-Aware Computing (PAC)}.\hskip 1em plus 0.5em minus 0.4em\relax IEEE,
  2018, pp. 19--30.

\bibitem{economic}
J.~M. Abowd and I.~M. Schmutte, ``An economic analysis of privacy protection
  and statistical accuracy as social choices,'' \emph{American Economic
  Review}, vol. 109, no.~1, pp. 171--202, 2019.

\bibitem{staircase}
Q.~Geng, P.~Kairouz, S.~Oh, and P.~Viswanath, ``The staircase mechanism in
  differential privacy,'' \emph{IEEE Journal of Selected Topics in Signal
  Processing}, vol.~9, no.~7, pp. 1176--1184, 2015.

\bibitem{ghosh2012universally}
A.~Ghosh, T.~Roughgarden, and M.~Sundararajan, ``Universally utility-maximizing
  privacy mechanisms,'' \emph{SIAM Journal on Computing}, vol.~41, no.~6, pp.
  1673--1693, 2012.

\bibitem{natashalaplace}
N.~Fernandes, A.~McIver, and C.~Morgan, ``The {Laplace} mechanism is optimal
  for differential privacy over continuous queries.'' in \emph{ACM/IEEE
  Symposium on Logic in Computer Science (LICS) (to appear)}, 2021.

\bibitem{individual}
J.~Soria-Comas, J.~Domingo-Ferrer, D.~Sánchez, and D.~Megías, ``Individual
  differential privacy: {A} utility-preserving formulation of differential
  privacy guarantees,'' \emph{{IEEE} Trans. Inf. Forensics Security}, vol.~12,
  no.~6, pp. 1418--1429, June 2017.

\bibitem{data_depenedent}
I.~Kotsogiannis, A.~Machanavajjhala, M.~Hay, and G.~Miklau, ``Pythia: Data
  dependent differentially private algorithm selection,'' 05 2017, pp.
  1323--1337.

\bibitem{personalized}
Z.~Jorgensen, T.~Yu, and G.~Cormode, ``Conservative or liberal? personalized
  differential privacy,'' in \emph{2015 IEEE 31St international conference on
  data engineering}.\hskip 1em plus 0.5em minus 0.4em\relax IEEE, 2015, pp.
  1023--1034.

\bibitem{RafnoDP2colorPaper}
R.~G.~L. D'Oliveira, M.~M{\'e}dard, and P.~Sadeghi, ``Differential privacy for
  binary functions via randomized graph colorings,'' in \emph{{IEEE} Int. Symp.
  Inf. Theory}, Melbourne, Victoria, Australia, July 2021, pp. 473--478.

\bibitem{dmetric}
K.~Chatzikokolakis, M.~E. Andr{\'e}s, N.~E. Bordenabe, and C.~Palamidessi,
  ``Broadening the scope of differential privacy using metrics,'' in
  \emph{International Symposium on Privacy Enhancing Technologies
  Symposium}.\hskip 1em plus 0.5em minus 0.4em\relax Springer, 2013, pp.
  82--102.

\end{thebibliography}


\end{document}